\documentclass[preprint2]{aastex}
\usepackage{graphicx}
\usepackage{color}
\usepackage{multirow}

\begin{document}

\title{On the resolution of a weak Fermi paradox}

\author{Osmanov Z.N.}
\affil{School of Physics, Free University of Tbilisi, 0183, Tbilisi,
Georgia\\
}

\begin{abstract}
In this paper we consider the Fermi's paradox (FP) and propose a possible resolution in the context of super-advanced alien civilizations. In this sense it can be regarded as to be the weak Fermi paradox (WFP). By assuming that superadvanced extraterrestrials (ET) exist and are capable of constructing a huge ring-like megastructure around a pulsar or a main sequence star (MSS) to consume its emitted energy, we consider the stability problem of the ring, studying the out of plane dynamics. It has been shown that for normal pulsars the oscillation timescale, thus the timescale of the spectral variability is of the order of several days, whereas for MSS it can vary from minutes up to hundreds of years. We have shown that by means of the high resolving power and sensitivity, the European Southern Observatory's Very Large Telescope can detect either cold ($T=300$ K) or hot ($T=4000$ K) megastructures throughout the whole galaxy even for very small oscillation amplitudes of rings. As it is shown the observed variability can have characteristics similar to those of some other variables, which can be considered as one of the resolutions of the WFP. 

\end{abstract}


\section{Introduction}

\begin{figure}
  \centering {\includegraphics[width=7cm]{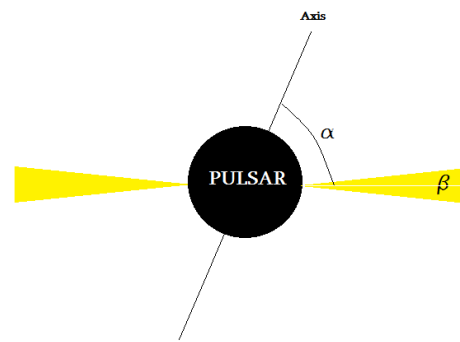}}
  \caption{Here we schematically show the pulsar, its axis of rotation,
  and two emission channels (yellow) with an opening angle $\beta$.}\label{fig1}
\end{figure}


\begin{figure}
  \centering {\includegraphics[width=7cm]{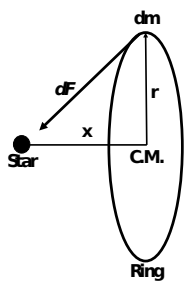}}
  \caption{Here we show the out-of-plane displaced ring with respect to the star (either a pulsar or a main sequence star).}\label{fig2}
\end{figure}

In 1950 discussing the problem of extraterrestrial intelligence the physicist Enrico Fermi asked "where are they?". The scientific background of this question and consequently the Fermi's Paradox (FP) is following: there are approximately $2\times 10^{11}$ stars in our local galaxy, if only a tiny fraction, $10^{-9}$, \citep{armstrong} of these worlds develop intelligence, there would be approximately $200$ ETs. On average, the terrestrial planets in the Milky Way are older than Earth by $(0.9-2.7)$ billion years \citep{age}. Therefore, if the intelligent beings exist, they should be more advanced than we are, but then the natural question arises: {\it where are they?}. If the answer to this question is not trivial (we do not see them because they do not exist), then the problem is very fundamental and serious.

There have been proposed a series of possible solutions to the FP. In particular, \cite{forest} suggests a solution 
based on an assumption that civilizations in the universe compete for limited resources and therefore, in the framework of this paradigm, the Universe is a dark forest, where any civilization is a hunter and any life-form exhibiting its existence must be terminated. In the context of communication by using the so-called nodes presented in \citep{nodes}, the author concludes that we are blind to the signals used by ETs because the corresponding beams are very narrow. Apart from the mentioned works, there is a number of other approaches to the problem of FP, which are analysed in detail \citep{silence} and it is argued that to resolve the problem one needs {\it a new spirit of open mindedness towards original, speculative and even crazy leaps.}

In 1960 Dyson has been proposed a very original idea in the framework of the Search for Extraterrestrial Intelligence (SETI). In particular, assuming that if an extraterrestrial intelligence exists and has reached 
a super technological level, capable of consuming the whole energy of their host star, it can build a spherical 
megastructure surrounding the central object \citep{dyson}. Theoretical reasonability of Dyson's approach was based on two values: our civilization's total power of energy consumption, $4\times 10^{19}$ergs s$^{-1}$ and the bolometric luminosity of the Sun-like stars, $4\times 10^{33}$ergs s$^{-1}$. Then, by assuming that $1\%$ of annual increase in industry is maintained, one can straightforwardly show that in approximately $3000$ years the civilization might reach the mentioned level of technology - Level-II in Kardashev's scale
\citep{kardashev}. Based on these assumptions \cite{dyson} theorises that for the purpose of harnessing the 
whole energy of the star, the civilization has to construct a thin spherical shell - Dyson sphere (DS) around this object. Then it is clear that the megastructure's surface located in the habitable zone (HZ) should be visible in the infrared spectral band and therefore might be detected by infrared telescopes.

In a historical perspective, the search for the DS candidates is relatively new challenge. The last two decades are somewhat prospective, because of several attempts to identify on the sky signatures of DSs \citep{jugaku,timofeev}. An interesting work has been done by \cite{carrigan}, who analysing 
the data of The Infrared Astronomical Satellite, has identified $16$ interesting candidates. A fascinating discovery was announced by \cite{tabby}: they have detected a variable object KIC8462852 (Tabby's star) with anomalously high flux dips, which cannot be explained by massive planet shading. Although, as it turned out, the anomalous behaviour of flux should be caused by exocomets \citep{tabbydust}, the discovery of the Tabby's star has stimulated a further study in this field and in a certain sense, it has revived the Dysonian SETI. Recently, by taking into account the observational data of the Gaia mission, \cite{gaia} emphasised peculiarity of the object 
TYC 6111-1162-1, which, according to the authors, deserves further investigation. The search for even more advanced, Level-III civilizations (capable of harnessing almost the whole energy of the galaxy \citep{kardashev})
has been performed by \cite{grif}, who scrutinly studied the data of the Wide-field Infrared Survey Explorer (WISE), and have shown that among $10^5$ galaxies $100$ deserve a certain attention. A special case of galaxies partially cloaked by DSs was considered by \cite{lacki} where the author has examined the corresponding observational pattern of such galaxies.

On the other hand, despite some attempts the SETI project in its broad sense (Dysonian SETI is only one particular branch) has failed and there is a serious scientific discussion about it \citep{melkikh,cirkovic2,cirkovic3,cirkovic1} where it is argued that conventional methods and approaches must be widened. In this context, recently the original idea has been developed by \cite{paper1,paper2} where the problem of Dysonian SETI was presented in a rather exotic way. In particular, the author has considered the possibility of colonising a nearby zone of a pulsar to utilise the whole radiated energy. Since the pulsars have narrow emission channels (see Fig.1), instead of a sphere, one should use a ring-like structure, requiring at least two orders of magnitude less material than for DSs \citep{paper1}. In \citep{paper2} the problem has been studied in the light of an observational pattern of these megastructures and their possible detection by ground based facilities. It is worth noting that pulsars as possible sites for advanced ET activities deserve more and more attention. In particular, \cite{beacon} has considered pulsars as possible beacons used by ETs for communication.

Another interesting feature, although for DSs has been considered in \citep{paper3} where we have studied the possible oscillation of the megastructure, leading to potentially detectable spectral variability \footnote{I am grateful to Prof. J. Wright for making aware of an incorrect conclusion in \citep{paper3}, emphasising that the fully complete DS will not lead to pressure difference. Although, not completely closed DS or a ring, will inevitably lead to the spectral variability.}. In the recent study by \cite{wright} it is shown that to maintain stability of the monolithic DS the elastic modulus of the material the megastructure is made of, should exceed the elastic modulus of the strongest material Graphen by many orders of magnitude, which indicates that monolithic spherical shells around stars are unrealistic. Therefore, in the present paper we consider the rings/incomplete DSs nearby pulsars and stars and study their variability in the context of the WFP.

The paper is organised in the following way: in Sec. 2, we develop a main approach, and in Sec. 3 
we summarize our results.

\section[]{Main approach}
 We do not consider the problem of a non-trivial connection of probability of the existence of intelligent life-forms and the number of biosignatures \citep{foucher}. In the framework of the paper we assume that if ETs have reached a level of super advanced civilization (Level-II), they might be able to "colonise" a nearby zone of either a pulsar or a star and construct a huge megastructure around them, which should be visible in the infrared or optical spectral band. 

If the DS is complete, it will be in the so-called neutral equilibrium. in particular, from the Gauss law, it is clear that the gravitational net force acting on the megastructure equals zero. The star's radiation in turn, is spherically symmetric, the momentum transferred by the emission is symmetric as well and therefore, the radiation pressure net force is zero. On the other hand, non-spherical structures might lead to non-zero net force difference and consequently to the spectral variability. We consider here an original idea of Niven rings and calculate the corresponding emission pattern.

\cite{stability} has considered a point mass and a thin solid ring to study non-linear
dynamics of the latter. The author examined two different regimes, the so-called in-plane and out-of-plane
motion of the ring. In the former case he has shown that the process is unstable. Studying 
the ring-like megastructure around pulsars \cite{paper1} has shown that pulsar's efficient radiation pressure cannot stabilise the process, but restoring the equilibrium position requires power, small compared to the total power received from the central object.

In Fig. 2 we schematically show the out-of-plane displaced ring. $x$ represents the displacement of the C.M. (the centre of mass) from the central object's position. The overall stability in out-of-plane motion is provided by the $x$ component of the gravitational force between the star (either pulsar, or a normal star) and the ring
\begin{equation}
\label{grav1} 
F_x = -G\frac{M_{st}Mx}{\left(R^2+x^2\right)^{3/2}},
\end{equation}
for small amplitude oscillations reducing to the following expression
\begin{equation}
\label{grav1} 
F_x \simeq -G\frac{M_{st}M}{R^2}x,
\end{equation}
where $G$ is the gravitational constant, $M_{st}$ represents the star's mass and $M$ is the mass of the ring. Then, one can straightforwardly show that the equation of motion is described by 
\begin{equation}
\label{eqm1} 
\frac{d^2\xi}{dt^2}+\omega^2\xi = 0,
\end{equation}
with $\xi\equiv x/R$ and $\omega = \sqrt{GM_{st}/R^3}$.

\subsection{Megastructures around pulsars} 

As a first example we examine pulsars, studying possible observational signatures

By taking into account that the total effective area of the ring is
$A_{_{ef}} = 8\pi R^2\sin\left(\beta/2\right)$, one can straightforwardly derive from energy balance 
the radius of the ring for a given black body temperature, $T$
$$R=
\left(\frac{L_p}{8\pi\sigma\sin\left(\beta/2\right)T^4}\right)^{1/2}\simeq $$
\begin{equation}
\label{R0} 
\simeq 0.038\times 
\left(\frac{L_p}{10^{30}ergs/s}\right)^{1/2}\times
\left(\frac{300K}{T}\right)^2 AU,
\end{equation}
where $L_p$ is the bolometric luminosity of a pulsar, normalised by the typical value of normal pulsars, $\sigma\approx 5.67\times 10^{-5}$erg/(cm$^2$K$^4$) is the
Stefan-Boltzmann's constant and temperature is normalised by the value corresponding to the HZ. We have also used the typical value of the opening angle, $\beta\simeq 32^o$ \citep{rudsuth}. As it is clear from Eq. (\ref{R}), the radius of the ring is much smaller than the radii of typical DSs $\sim 1$ AU. Therefore, an amount of material required for constructing the ring will be much less than for building the standard DS. On the other hand, one can assume that a reasonable tendency should be to decrease the total mass of the megastructure. In OB, we have examined Graphene as a typical example of a super strong material having a very high melting temperature $4510$ K \citep{graphene} and therefore, it can be located much closer to the host star than other materials. In fact, we assume that if our civilization can produce super light and super strong materials, the same job will not be a problem for alien Level-II civilizations. From Eq. (\ref{R0}) one can straightforwardly show that for $T = 4000$ K the radii of the rings will be of the order of $2\times 10^{-4}$ AU, which is by four orders of magnitude less than normal DS radii. In this case the megastructure's spectrum will 
peak at the wavelength, $725$ nm, which in turn, means that it can be seen not only in far infrared, but also in
the optical band. In this paper we consider only two typical temperatures: the normal temperature ($300$K) and the maximum possible temperature ($\sim 4000$K) as more or less two extreme cases.

On the other hand, as we have discussed in OB, some sector of the inner free space of the ring should be maintained relatively cold. If the megastructure is used for habitation, the temperature should be around $300$ K and for providing high performance computing processes \citep{sandberg} the temperature should be of the order of $10$ K. By taking into account that the typical values of modern cooling engines have the coefficient of performance (COP) of the order of $5$, the engine to compensate the flux to the cold area, must process the energy (normalised by the total luminosity of the pulsar) to the hot reservoir,

$$P_e\simeq\frac{1}{COP}\frac{\kappa S}{L_p }\frac{\Delta T}{h}\simeq $$
\begin{equation}
\label{cool} 
\simeq 0.01\times 
\frac{\Delta T/h}{37 K cm^{-1}}\times\frac{\kappa}{2.5\times 10^8 erg (cmK)^{-1}},
\end{equation}
where we assumed $\Delta T = (4000-300) K = 3700 K$, $h\simeq 100$ cm, $\kappa$ is the thermal conductivity normalised by the corresponding value for Graphene \citep{graphene1}, $S$ is the area inside the free space requiring cooling and we have considered the value equal to the total surface area of the Earth, $5.1\times 10^8$km$^2$. In case the megastructure is used for high performance computing, one can see from the above equation that $P_e$ will be still much less than the pulsar's luminosity. Therefore, from the point of view of laws of physics the process of cooling is quite realistic.


By combining Eqs. ({\ref{eqm1},\ref{R0}}) the oscillation period writes as
$$P_p = 2\pi\sqrt{\frac{R^3}{GM_p}}\simeq3\times \left(\frac{1.5\times M_{s}}{M_p}\right)^{1/2}\times$$
\begin{equation}
\label{per} 
 \times\left(\frac{L_p}{10^{30}ergs/s}\right)^{3/4}
\times\left(\frac{300 K}{T}\right)^3 days,
\end{equation}
where $M_p=1.5\times M_{\odot}$ represents the pulsar's mass and $M_{\odot}\simeq 2\times 10^{33}$g is the Solar mass.

Such a periodic motion will inevitably lead to a variable character of an observational pattern. In particular, by means of the Doppler effect, the observed wavelength is given by \citep{carroll}
\begin{equation}
\label{dopp1} 
\lambda = \lambda_s\frac{1+v\cos\theta/c}{\sqrt{1-v^2/c^2}},
\end{equation}
where by $\lambda_s$ we represent the original wavelength emitted by the ring, $\theta$ is an angle of velocity direction measured in an observer's frame of reference, $v$ denotes the corresponding velocity and $c$ is the speed of light. It is clear that oscillation is non-relativistic, which for the maximum wavelength difference leads to
\begin{equation}
\label{dopp2} 
\Delta\lambda\simeq \lambda_s\frac{2v_m\cos\theta}{c},
\end{equation}
where multiplier $2$ comes from motion of the ring in different directions, $v_m = \omega A$ denotes the maximum velocity and $A =\chi R$ is the amplitude of small oscillations ($\chi << 1$). A technical problem for observing such megastructures is to achieve high values of the resolving power, $RP$, of modern spectrographs. As we have already mentioned, the spectral band, in which these rings might be detected, is either in infrared (for $T = 300$ K) or far infrared-optical interval (for $T = 4000$ K). Up to now the European Southern Observatory's (ESO) Very Large Telescope (VLT) has the highest resolving power $RP\equiv\Delta\lambda/\lambda = 25000$ for $\lambda \approx 9.6\mu$m (which corresponds to the blackbody radiation with $T=300$K) \footnote{www.eso.org/public/teles-instr/paranal-observatory/vlt/}. On the other hand, it is clear that spectral variations will be detected if the following relation $\Delta\lambda/\lambda_s\leq RP$ is satisfied. This imposes a certain constraint on pulsar luminosities 
which potentially might be observed
%
$$\chi\geq \frac{c}{RP\;T\cos\theta}\times\left(\frac{L_p}{64\pi\beta\sigma G^2M_p^2}\right)^{1/4}\simeq 0.035\times$$
\begin{equation}
\label{cond1} 
\times \frac{1}{\cos\theta}\times\frac{2500}{RP}\times\frac{300 K}{T}\times\left(\frac{L_p}{10^{30}erg/sec}\right)^{1/4},
\end{equation}
By taking into account that ring's luminosity behaves with pulsar's rotation period as $L\propto P^{-3}$ \citep{paper1} and since the upper limit of the average value of periods is of the order of $2$sec \citep{taylor}, one can estimate the minimum value of $\chi_{m}\sim 3\times10^{-2}$, which will still allow to detect the variability. 

For hot rings ($T=4000$K) the intensity peaks at $\lambda\approx 725$nm and the corresponding limit of the resolving power of the VLT facility is by one order of magnitude higher, $RP = 190000$, which reduces the minimum value of the spatial amplitude by one order of magnitude, $\chi_m\simeq 3\times 10^{-4}$. Therefore, for hot rings the spectral variation might be detected for even smaller amplitude scales.

Generally speaking, if the observational pattern of the megastructure is characterised by absorption lines, then by using the so-called radial velocity method \citep{radvel} it is possible to measure radial velocity by precision of the order of $\upsilon_m\sim 1$m/s. For this case, one can obtain the constraint on potentially detected luminosities
$$\chi\geq \frac{\upsilon_m}{T}\times\left(\frac{L_p}{4\pi\beta\sigma G^2M_p^2}\right)^{1/4}\simeq 6\times 10^{-6}\times$$
\begin{equation}
\label{cond2} 
\times\frac{\upsilon_m}{1m/sec}\times\frac{300 K}{T}\times\left(\frac{L_p}{10^{30}erg/sec}\right)^{1/4},
\end{equation}
which for $T = 300$K and the same class of pulsars leads to the minimum value of the oscillation amplitude $\chi_m\sim 5\times 10^{-6}$ and for $T = 4000$K the corresponding value is of the order of $\chi_m\sim 4\times 10^{-7}$.

\cite{stability} has shown that the ring is unstable with in-plane displacements with the corresponding timescale, $\tau = \sqrt{2R^3/(GM_{c})}$, where $R$ is the ring radius and $M_c$ is the central object's mass. This in turn, means, that the characteristic velocity of the ring becomes of the order of $\upsilon\simeq\delta R/\tau$ where $\delta R$ is the initial displacement of the ring from the equilibrium position. On the other hand, in the Lunar laser ranging experiment \footnote{The corresponding data is available from the
Paris Observatory Lunar Analysis Center:
http://polac.obspm.fr/llrdatae.html} the distance is measured with the precision of the order of $\eta\simeq 10^{-10}$. Then, if one assumes that the precision of the super advanced civilization controlling the distance is not less than $\eta$, one can straightforwardly obtain $\upsilon\simeq\eta\sqrt{GM_c/(2R_c)}$, which for typical pulsars leads to a value in the interval $10^{-3}-10^{-2}$cm/sec, being beyond the sensitivity of the modern facilities. It is worth noting that it would be possible to detect such megastructures if the measurement precision will be reduced by the factor of $10^{4-5}$, which although is possible but less probable: if our civilization can measure distances with the precision $10^{-10}$, one can naturally assume that a super-advanced alien civilization might have even higher precision measurements.

\cite{paper2} has demonstrated the possibility of the VLT to distinguish the structure of the ring in the HZ and it has been shown that in the nearby zone of the Solar system (on a distance $\sim 0.2$kpc) for approximately 64 pulsars the megastructures might be observed in detail. By using the method of spectral variability, not only nearby pulsars, but very distant objects, located almost at the edge of our galaxy, might be monitored as well. In particular, the flux sensitivity of the VLT instruments is approximately $Flux \simeq 3\times 10^{-18}$ ergs s$^{-1}$, which means that the maximum distance from which the radiation of the ring still might be detectable is given by
\begin{equation}
\label{dist} 
D_{max}\simeq\sqrt{\frac{L}{4\pi\times Flux}}\simeq 50\times\sqrt{\frac{L}{10^{30}ergs/s}}kpc,
\end{equation}
which is of the same order of magnitude as diameter of Milky Way.

\subsection{Megastructures around MSS}

In this subsection we examine the megastructures around MSS and outline the major resutls presented in \citep{paper4}. We consider the ring like constructions. Then the radius of the ring for the Solar-type stars is given by
$$R_{st}=
\left(\frac{L_{st}}{8\pi\sigma T^4}\right)^{1/2}\simeq $$
\begin{equation}
\label{R} 
\simeq 1.22\times 
\left(\frac{L_{st}}{L_{s}}\right)^{1/2}\times
\left(\frac{300K}{T}\right)^2 AU,
\end{equation}
where the stellar bolometric luminosity, $L_{st}$, is normalised by the solar luminosity, $L_{\odot}\simeq 3.83\times 10^{33}$ergs s$^{-1}$. For the hot Dyson megastructure ($T = 4000$K) the corresponding scale is of the order of $0.007$AU.

Consequently, likewise the previous case, the variability timescale is expressed as follows
$$P_{st} = 2\pi\sqrt{\frac{R_{st}^3}{GM_{st}}}\simeq 490\times \left(\frac{M_{s}}{M_{st}}\right)^{1/2}\times$$
\begin{equation}
\label{per1} 
 \times\left(\frac{L_{st}}{L_{s}}\right)^{3/4}
\times\left(\frac{300 K}{T}\right)^3 days.
\end{equation}
%

We consider a wide range of main sequence stars with masses $0.1 M_{\odot} < M_{st}<120 M_{\odot}$, characterized by the following approximate luminosity-mass relation \citep{lummass}
\begin{equation}
\label{masslum} 
 L_{st}\simeq\alpha_1\times L_{s}\times\left(\frac{M_{st}}{M_{s}}\right)^{\alpha_2},
\end{equation}
where $\alpha_1\simeq 1.03$ and $\alpha_2\simeq 3.42$. In Table 1 we show the variability time-scales for different classes of stars. As it is clear, the time-scale might become anomalously short, of the order of minutes and hours.

By taking the aforementioned equations into account, one can obtain the constraint on dimensionless spatial amplitude

\begin{table}
\begin{tabular}{ |l|l|l| }
\hline
\multicolumn{3}{ |c| }{Variability Time-Scale for MSS} \\
\hline
Class & $T = 300$K & $T = 4000$K \\ \hline
O & Hundreds of years & Months - Years \\  \hline
B & Years - Hundreds of years & Day - Months \\ \hline
A & Years & Hours - Day \\ \hline
F & Years & Hours \\ \hline
G & Years & Hours \\ \hline
K & Months - Year & Hours \\ \hline
M & Days - Months & Minutes - Hour \\ \hline
\hline

\end{tabular}
\caption{Here we present the variability time-scales for Main Sequence Stars.}
\end{table}

$$\chi\geq \frac{c}{2Rp\;T\cos\theta}\times\left(\frac{L}{8\pi\sigma G^2M^2}\right)^{1/4}\simeq 0.22\times$$
\begin{equation}
\label{cond2} 
\times \frac{1}{\cos\theta}\times\frac{2500}{RP}\times\frac{300 K}{T}\times\left(\frac{M_{st}}{M_{s}}\right)^{\frac{\alpha_2-2}{4}}.
\end{equation}
It is evident that for cold Dyson megastructures, more promising objects are M-type stars, but even for them the minimum value of the dimensionless amplitude is of the order of $0.1$. For hot rings, the corresponding estimate leads to the value, $0.001$. (See Table 2.)

The radial velocity method still might be very efficient. In particular, the corresponding constraint in this case writes as
$$\chi\geq \frac{\upsilon_m}{T}\times\left(\frac{L}{8\pi\sigma G^2M^2}\right)^{1/4}\simeq 3.72\times 10^{-5}\times$$
\begin{equation}
\label{cond3} 
\times \frac{\upsilon_m}{1m/sec}\times\frac{300 K}{T}\times\left(\frac{M_{st}}{M_{s}}\right)^{\frac{\alpha_2-2}{4}},
\end{equation}
which even for Solar type stars and cold megastructures leads to very small value $\chi_{min}\sim 3.72\times 10^{-5}$. The hot Dyson rings with $T=4000$K lead to even smaller lower threshold, $2.79\times 10^{-6}$ (See Table 3).

Considering the in-plane instability of the ring for the wide range of stars, $0.1 M_{\odot} < M_{st}<120 M_{\odot}$ and by taking into account Eqs. (\ref{R}-\ref{masslum}) one can straightforwardly show that the velocity amplitude of in-plane oscillations are of the order of $10^{-5}-10^{-4}$, which is even smaller than in case of pulsars, implying that the modern facilities are not sensitive enough to detect the corresponding motion of megastructures.

\begin{table}
\begin{tabular}{ |l|l|l| }
\hline
\multicolumn{3}{ |c| }{Spectral variability method: $\chi$  for MSS} \\
\hline
Class & $T = 300$K & $T = 4000$K \\ \hline
O &  $>0.59$ & $\sim (5.9 - 12)\times 10^{-3}$ \\
  \hline

B & $\sim 0.29 - 0.59$ & $\sim (2.9 - 5.9)\times 10^{-3}$ \\ \hline
A & $\sim 0.24 - 0.29$ & $\sim (2.4 - 2.9)\times 10^{-3}$  \\ \hline
F & $\sim 0.22 - 0.24$ & $\sim (2.2 - 2.4)\times 10^{-3}$ \\ \hline
G & $\sim 0.2 - 0.22$ & $\sim (2 - 2.2)\times 10^{-3}$  \\ \hline
K & $\sim 0.16 - 0.2$ & $\sim (1.6 - 2)\times 10^{-3}$ \\ \hline
M & $\sim 0.1 - 0.16$ & $\sim (9 - 16)\times 10^{-4}$ \\ \hline
\hline

\end{tabular}
\caption{Here we show the values of $\chi$ for Main Sequence Stars derived in the framework of the spectral variability method.}
\end{table}

We analyse the obtained results separately for pulsars and stars. As a first example we consider pulsars. From Eq. (\ref{per}) it is evident that depending on temperature the timescales of variability might be several days or even less. On the other hand, modern observations in mid-infrared spectrum reveal the similar timescale variability from some astrophysical objects \citep{mid1,mid2,mid3}. {\it Therefore, if the spectral features of the rings around pulsars are indistinguishable from conventional astrophysical objects, the former might be "hidden"} and only very high spatial resolution can distinguish the internal structure of an observed object. At this moment the highest resolution, provided by the ESO's VLT instruments guarantee monitoring of only $64$ pulsars in the nearby zone of the Solar system.

By analysing MSS, from Eq. (\ref{per1}) one can conclude that in case of cold Dyson megastructures ($T = 300$K) the corresponding variability timescale might vary from days to almost a year (K and M stars), up to several years (G, F and A stars) and hundreds of years or higher (B and O stars). To identify Dyson megastructure candidates by detecting variability one should analyse the overall spectral picture. In particular, the long period stars belong to the spectral class F, M, S or C, but F stars cannot be considered as candidates since they are very hot ($T=7000$K). M stars have temperatures corresponding to hot DSs $(2400-4700)$K, but are unrealistically luminous like S and C stars. On the other hand, as it is clear from Eq. (\ref{per1}), for stars with $M>7M_{\odot}$ the variability timescale becomes more than $100$ years and therefore, to discover them the observation time should be that large. This might be one of the resolutions of the FP - we do not have enough time to detect such objects, the technological artefacts of highly advanced civlizations.

Studying the hot megastructures, one can see that starting from M up to A type stars the timescale varies from several minutes to approximately a day. For B type stars the corresponding value varies from a day up to several months and for O type stars the variability timescale interval might be from months to years. One should emphasize that the ZZ Ceti stars are characterised by the variability timescale of the order of several minutes \citep{carroll}, but these stars have extremely high temperatures $10^4$K. The stars of spectral class B are characterised by variability timescales of the order of several hours, but still the temperature is very high. The similar inconsistency occurs with variability timescales from days to a year. Therefore, in order to find real candidates one should focus on anomalous spectral behaviour of variables.

In the framework of the paper we have assumed that the technosignatures of Level-I civilization should not be easily detectable because of the relatively small energy consumption. The existence of Level-III civilization is possible, but less probable, than Level-II, therefore, the paper is focused on the latter. 

We have analysed the observational features of megastructures nearby stars and pulsars. In case of stars, if the megastructures are complete DSs then as we have already discussed, it will not be characterised by variability and therefore, cannot be detected. On the other hand, the incomplete megastructures - rings, will exhibit the strong variability in out of plane dynamics, but the emission characteristics of such rings, will be somehow "hidden" in the overall emission background of the sky and to discover a candidate one has to search for technosignatures by searching for anomalous variable "stars". It is also clear that to consume the whole energy of a star, one should construct the complete DS, requiring enormous material. Therefore, one should naturally assume the possibility of colonising pulsars - the advantage could be the usage of much less material. However, it is worth noting that the construction of megastructures around pulsars requires colonization and thus - migration. Sumarizing, one can conclude that the technosignatures are either non-detectable (complete DSs around stars), or hidden in the sky's backgrount (rings nearby stars or pulsars).

\begin{table}
\begin{tabular}{ |l|l|l| }
\hline
\multicolumn{3}{ |c| }{Radial velocity method: $\chi$  for MSS} \\
\hline
Class & $T = 300$K & $T = 4000$K \\ \hline
O &  $\sim(1 - 2)\times 10^{-4}$ & $\sim (7.5 - 15)\times 10^{-6}$ \\ \hline
B & $\sim(0.49 - 1)\times 10^{-4}$ & $\sim (3.6 - 7.5)\times 10^{-6}$ \\ \hline
A & $\sim(0.42 - 0.49)\times 10^{-4}$ & $\sim (3.2 - 3.6)\times 10^{-6}$  \\ \hline
F & $\sim(0.38 - 0.42)\times 10^{-4}$ &  $\sim (2.8 - 3.2)\times 10^{-6}$ \\ \hline
G & $\sim(0.34 - 0.38)\times 10^{-4}$ & $\sim (2.6 - 2.8)\times 10^{-6}$   \\ \hline
K & $\sim(0.28 - 0.34)\times 10^{-4}$ & $\sim (2.1 - 2.6)\times 10^{-6}$ \\ \hline
M &  $\sim(0.16 - 0.28)\times 10^{-4}$ & $\sim (1.2 - 2.1)\times 10^{-6}$ \\ \hline
\hline

\end{tabular}
\caption{Here we show the values of $\chi$ for Main Sequence Stars derived in the framework of the radial velocity method.}
\end{table}

\section{Conclusion}

By considering the out-of-plane dynamics of the ring we have estimated the oscillation period of the megastructure. For reasonable parameters of stars and pulsars it has been shown that the typical timescale 
of oscillations and thus the timescales of variability might vary from minutes (hot megastructures around M stars or pulsars) to days (cold rings around M stars or pulsars) up to hundreds of years (cold megastructures around B and O type stars).

Two methods of detecting the technosignatures have been examined: the spectral variability and radial velocity method. We have studied as relatively cold ($T=300$ K) as well as hot ($T=4000$ K) megastructures. It was shown that by ESO's VLT instruments the variable character of spectra can be detected even for small oscillations $\chi\sim\left(0.001-0.01\right)$ (for pulsars) and $\chi\sim\left(0.001-0.1\right)$ (for stars). By considering the radial velocity method the results are following: $\chi\sim\left(10^{-7}-10^{-6}\right)$ (for pulsars) and $\chi\sim\left(10^{-6}-10^{-4}\right)$ (for stars).

It has been shown that rings might be spectrally indistinguishable from some variables - thus, they might be hidden, which in turn is one of the resolutions of the WFP. On the other hand, a superadvanced civilization might control the oscillation amplitude - making it extremely small, shifting the variability in an undetected "area". In any case, even if the megastructures are detectable, they should reveal anomalous character of their variability, which is the main reason why we cannot see them: we look for aliens in wrong "areas".


\section*{Acknowledgments}
The author is grateful to Prof. J. Wright, emphasising that the fully complete DS will not lead to pressure difference. 

\end{document}